\documentstyle[aps,prl,twocolumn,epsf,floats,12pt]{revtex}

\draft
\begin{document}
\title{Composite Fermions and the Fractional Quantum Hall Effect:
       Essential Role of the Pseudopotential\vspace*{-1ex}}
\author{
   \underline{J. J. Quinn} and A. W\'ojs}
\address{\footnotesize\sl
   University of Tennessee, Knoxville, Tennessee 37996, USA\\[2ex]}
\address{
   \footnotesize\rm\parbox{6.5in}{
   The mean field (MF) composite Fermion (CF) picture successfully 
   predicts the band of low lying angular momentum multiplets of 
   fractional quantum Hall systems for any value of the magnetic field.
   This success cannot be attributed to a cancellation between Coulomb 
   and Chern--Simons interactions between fluctuations beyond the mean 
   field.
   It results instead from the short range behavior of the Coulomb 
   pseudopotential in the lowest Landau level (LL).
   The class of pseudopotentials for which the MFCF picture is successful
   can be defined, and used to explain the success or failure of the 
   picture in different cases (e.g. excited LL's,  charged magneto-excitons, 
   and Laughlin quasiparticles in a CF hierarchy picture).}\\[-3ex]}
\maketitle
\paragraph*{Introduction.}
The MFCF picture\cite{jain,chen} does remarkably well in predicting the
band of angular momentum ($L$) multiplets that form the low energy sector 
of a 2D electron system in a strong magnetic field $B$.
A Laughlin\cite{laughlin} incompressible $L=0$ ground state of an $N$
electron system occurs when the magnetic monopole (which produces the 
radial magnetic field at the surface of the Haldane\cite{haldane1} 
sphere) has strength $2S_m=m(N-1)$, where $m$ in an odd integer.
\begin{figure}[t]
\epsfxsize=3.1in
\epsffile{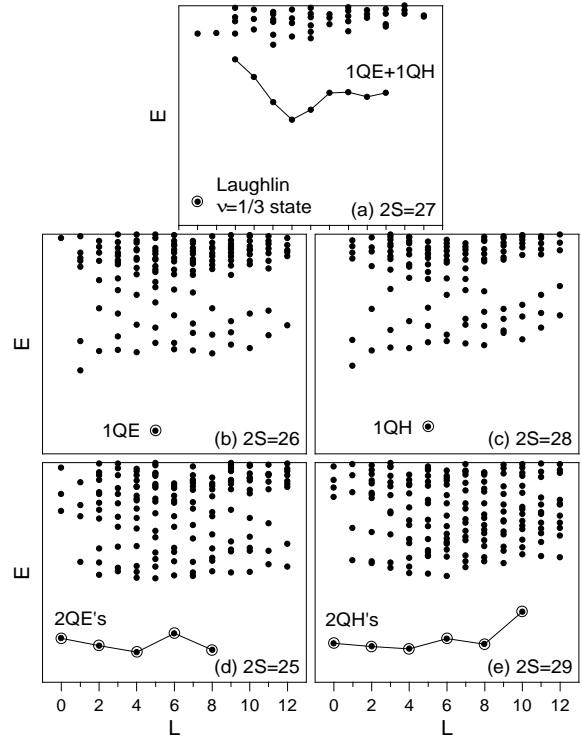}
\caption{
   Energy spectra of ten electrons in the lowest LL at $25\le2S\le29$. 
   Open circles mark lowest energy bands with fewest CF QP's.}
\label{fig1}
\end{figure}
For $2S$ different from $2S_m$ there will be $|2S-2S_m|$ quasiparticles
(QP's).
This is illustrated in Fig.~\ref{fig1}, which displays the energy spectra
of ten electrons on a Haldane sphere at monopole strength $25\le2S\le29$.
Frame (a) shows the Laughlin incompressible ground state at $L=0$.
Frames (b) and (c) show states containing a single quasielectron 
QE (a) and quasihole QH (b) at $L=5$.
In frames (d) and (e) the two QP states form the low energy bands.
In the MFCF picture, the effective monopole strength $2S^*$ is given 
by $2S^*=2S-2p(N-1)$, where $p$ is an integer.
$S^*$ is the angular momentum $l_0^*$ of a MF CF in the lowest CF Landau
level.
At $2S=27$ (with $p=1$), $l_0^*=9/2$ and the lowest shell accommodates
$2l_0^*+1=10$ CF's, so that the shell is filled giving $L=0$.
At $2S=27\pm1$ there will be one CF QHwith $l_{\rm QH}=5$ or one CF QE 
with $l_{\rm QE}=5$, giving $L=5$.
At $2S=27\pm2$ there will be two CF QH each with $l_{\rm QH}=11/2$ giving 
$L=0$, 2, 4, 6, 8, 10, or two CF QE each with $l_{\rm QE}=9/2$ giving 
$L=0$, 2, 4, 6, 8.

It is quite remarkable that the MFCF picture works so well since its 
energy scale is $\hbar\omega_c^*=(2p+1)^{-1}\hbar\omega_c\propto B$,
in contrast to the scale of the Coulomb interaction $e^2/\lambda\propto
\sqrt{B}$, where $\lambda$ is the magnetic length.
The energy values obtained in the MFCF picture are totally incorrect,
but the structure of the low energy spectrum (which multiplets form
the lowest lying band) is correct.
As first suggested by Haldane\cite{haldane2}, this is a result of the
behavior of the pseudopotential $V(L')$ (interaction energy of a pair 
of electrons vs.\ pair angular momentum) in the lowest LL.

\begin{figure}[t]
\epsfxsize=3.1in
\epsffile{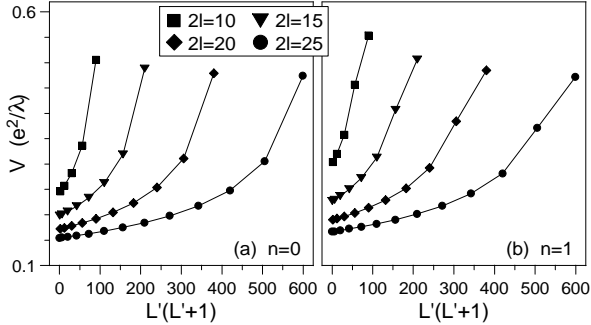}
\caption{
   Pseudopotentials $V$ of the Coulomb interaction in the lowest (a), 
   and first excited LL (b) as a function of squared pair angular 
   momentum $L'(L'+1)$ for different values of $l=S+n$.}
\label{fig2}
\end{figure}

\paragraph*{Pseudopotential.}
In Fig.~\ref{fig2} we plot $V(L')$ vs. $L'(L'+1)$ for the lowest ($n=0$) 
and first excited ($n=1$) LL for different values of $2l$\cite{wojs1}.
Note that for $n=0$ $V(L')$ rises more steeply than linearly with 
increasing $L'$ at all values of $L'$, but for $n=1$ this is not true
at the highest allowed values of $L'$.

A useful operator identity\cite{wojs2} relates the total angular momentum
$\hat{L}=\sum_i\hat{l}_i$ to the sum over all pairs of the pair angular
momentum $\hat{L}_{ij}=\hat{l}_i+\hat{l}_j$,
\begin{equation}
   \sum_{i<j} \hat{L}_{ij}^2 = \hat{L}^2 + N(N-2)\;\hat{l}^2.
\label{eq1}
\end{equation}
Here, each Fermion has angular momentum $l$, so that $\hat{l}^2$ has the
eigenvalue $l(l+1)$.
From Eq.~(\ref{eq1}) it is not difficult to show that for a ``harmonic''
pseudopotential defined by $V_H(L')=A+B\,L'(L'+1)$, the energy 
$E_\alpha(L)$ of the $\alpha$th multiplet with total angular momentum 
$L$ would be independent of $\alpha$, and that $E(L)$ would be of the 
form $a+b\,L(L+1)$\cite{wojs2}.
Because the actual pseudopotential is different from $V_H(L')$, the
degeneracy of the multiplets $\alpha$ of the same $L$ is lifted.

For a pseudopotential (which we will refer to as a short range, SR,
potential) that rises more quickly with $L'$ than $V_H(L')$, the lowest 
energy multiplets must, to the extent that it is possible, avoid having 
pair amplitude (or coefficient of fractional parentage \cite{shalit}) 
from the largest values of $L'$.
For $V_H(L')$ the lowest angular momentum states have the lowest energy.
However, the difference $\Delta V(L')=V(L')-V_H(L')$ lifts the 
degeneracy of multiplets having the same value of $L$.
If some low value of $L$  has a very large number $N_L$ of multiplets,
$\Delta V(L')$ can push the lowest multiplet at that $L$ value to a lower
energy than any multiplet of a neighboring smaller $L$ value for which
$N_L$ is much smaller.

\begin{figure}[t]
\epsfxsize=3.1in
\epsffile{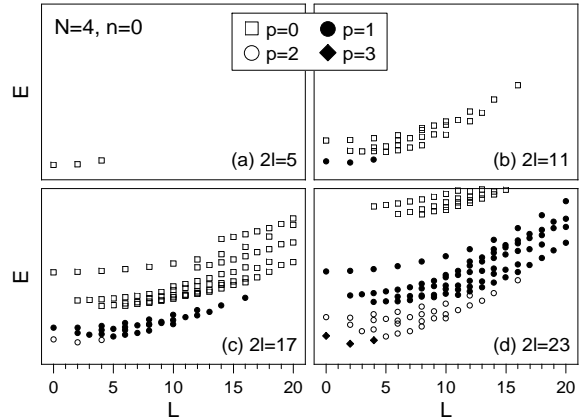}
\caption{
   Energy spectra of four electrons in the lowest LL.
   Different symbols mark subspaces ${\cal H}_p$ for $p=0$, 1, 2, and 3.}
\label{fig3}
\end{figure}

\paragraph*{Energy Spectra of SR Pseudopotential.}
Fig.~\ref{fig3} displays some very informative results\cite{wojs2} for
a simple four particle system at different values of the single particle
angular momentum $l$ (which differs by $\Delta l=p(N-1)$, $p=1$, 2, \dots).
Note that the set of multiplets at $l-p(N-1)$ is always the subset of the
multiplets at $l$.
The SR pseudopotential appears to have the property that its Hilbert 
space ${\cal H}$ splits into subspaces ${\cal H}_p$ containing states
with no parentage from pair angular momentum $L'=2(l-p)-1$.
${\cal H}_0$ is the entire space, ${\cal H}_1$ is the subspace that avoids 
$L'_{\rm MAX}=2l-1$, ${\cal H}_2$ avoids $L=2l-1$ and $2l-3$, etc.
Since the interaction energy in each subspace ${\cal H}_p$ is measured
on the scale $V(L'=2(l-p)-1)$, the spectrum splits into bands with gaps
between bands associated with the differences in appropriate 
pseudopotential coefficients.
The largest gap is always between the zeroth and first band, the next 
largest between the first and second, etc.
Note that the subset of multiplets at $l'=l-p(N-1)$ is exactly the 
subset chosen by the MFCF picture.
In addition, at the Jain values $\nu=n(1+2pn)^{-1}$, where $n=1$, 2, 
\dots, there is only a single multiplet at $L=0$ in the ``lowest subset''
for an appropriate value of $p$.

These ideas can be made more formal by using the algebra of angular 
momentum addition and the ``coefficients of fractional parentage''
familiar to atomic and nuclear physicists.
The conclusions are quite clear.
There is really only one energy scale, that of the Coulomb interaction
$e^2/\lambda$.
Laughlin states occur when the fractional parentage for electrons
(or holes) allows avoidance of the pseudopotential $V(2(l-p)-1)$ for
$p=0$, 1, \dots.
Jain states occur when the fractional parentage of the appropriate
$V(2(l-p)-1)$ is much smaller (but not zero) for $L=0$ than for other 
allowed multiplets.
The MFCF picture works only if $V(L')$ is a SR potential that rises like
$[L'(L'+1)]^\beta$ with $\beta>1$\cite{wojs2}.

\paragraph*{Other Pseudopotentials.}
For the $n=1$ and higher LL's, $V(L')$ is not SR for all values of $L'$.
For $n=1$, $V(L')$ is essentially harmonic at $L'=L'_{\rm MAX}$, and for
$n>1$ it is subharmonic at the largest values of $L'$.
Therefore, even if ground states at filling factors like $\nu=2+1/3$ 
have $L=0$, they are not Laughlin type incompressible states which
avoid pair angular momentum $L'_{\rm MAX}=2l-1$.

A CF hierarchy scheme was proposed by Sitko et al.\cite{sitko} in which
the CF transformation was reapplied to QP's in partially filled shells.
The application of the MFCF approximation was found to work in some cases
but not in others.
Some idea of when the MFCF approximation is valid can be obtained from
looking at the 2QE and 2QH states in Fig.~\ref{fig1}.
The QH pseudopotential is SR at $L=10$, but not at $L=8$.
The QE pseudopotential is certainly not SR at $L=8$, but at $L=6$ it 
might be.
This suggests that Laughlin states will be formed by QH's of the $\nu=1/3$ 
state at $\nu_{\rm QH}=1/3$ and by QE's of the $\nu=1/3$ state at 
$\nu_{\rm QE}=1$, explaining the strong FQHE of the underlying electron 
system at the Jain $\nu=2/7$ and 2/5 filling factors.
In contrast, no FQHE at $\nu_{\rm QH}=1/5$ ($\nu=4/13$ electron filling)
or $\nu_{\rm QE}=1/3$ ($\nu=4/11$) would be expected because the QP
pseudopotentials are not SR at these values.

A final interesting example is that of a multi-component plasma of 
electrons and one or more negatively charged excitonic ions $X_k^-$
(a bound state of $k$ neutral excitons and an electron) formed in 
an electron-hole system.
These excitonic ions are long lived Fermions with LL structure\cite{wojs3}.
The angular momentum of an $X_k^-$ on a Haldane sphere is $l_k=S-k$.
The pseudopotentials describing the interactions of $X_k^-$ ions with
electrons and with one another can be shown to be SR.
In fact, $V_{AB}(L')$, where $A$ or $B$ or both are composite particles
has a ``hard core'' for which one or more of the largest values of 
$V_{AB}(L')$ are effectively infinite.

\begin{figure}[t]
\epsfxsize=3.1in
\epsffile{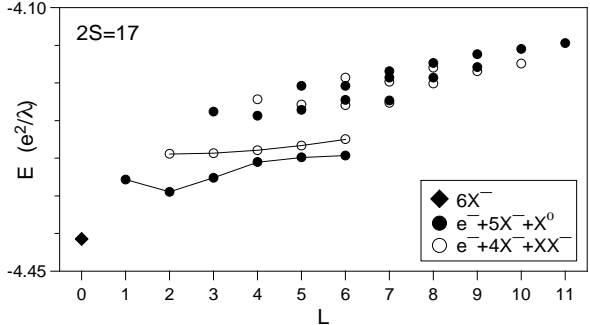}
\caption{
   Low energy spectra of different charge configurations of 
   the $12e+6h$ system on a Haldane sphere at $2S=17$.}
\label{fig4}
\end{figure}

The following configurations of ions have low energy in the twelve 
electron--six hole system at $2S=17$.
The $6X^-$ configuration (i) has the maximum total binding energy 
$\varepsilon_{\rm i}$.
Other expected low lying bands are:
(ii) $e^-+5X^-+X^0$ with $\varepsilon_{\rm ii}$ and 
(iii) $e^-+4X^-+X_2^-$ with $\varepsilon_{\rm iii}$.
Here, $\varepsilon_{\rm i}>\varepsilon_{\rm ii}>\varepsilon_{\rm iii}$ 
are all known.
Although we are unable to perform an exact diagonalization for this
system in terms individual electrons and holes, we can use appropriate 
pseudopotentials and binding energies to obtain the low lying states in 
the spectrum.
The results are presented in Fig.~\ref{fig4}.
There is only one $6X^-$ state (the $L=0$ Laughlin $\nu_{X^-}=1/3$ 
state\cite{laughlin}) and two bands of states in each of groupings 
(ii) and (iii).
A gap of 0.0626~$e^2/\lambda$ separates the $L=0$ ground state from 
the lowest excited state.

\paragraph*{Generalized Composite Fermion Picture.}
In order to understand the numerical results obtained in Fig.~\ref{fig4},
it is useful to introduce a generalized CF picture by attaching to each 
particle fictitious flux tubes carrying an integral number of flux quanta 
$\phi_0$.
In the multi-component system\cite{wojs4}, each $a$-particle carries flux 
$(m_{aa}-1)\phi_0$ that couples only to charges on all other $a$-particles 
and fluxes $m_{ab}\phi_0$ that couple only to charges on all $b$-particles,
where $a$ and $b$ are any of the types of Fermions.
The effective monopole strength seen by a CF of type $a$ (CF-$a$) is
$2S_a^*=2S-\sum_b(m_{ab}-\delta_{ab})(N_b-\delta_{ab})$.
For different multi-component systems we expect generalized Laughlin 
incompressible states when all the hard core pseudopotentials are avoided 
and CF's of each kind fill completely an integral number of their CF 
shells (e.g. $N_a=2l_a^*+1$ for the lowest shell).
In other cases, the low lying multiplets are expected to contain different 
kinds of quasiparticles (QP-$A$, QP-$B$, \dots) or quasiholes (QH-$A$, 
QH-$B$, \dots) in neighboring filled shells.
Our multi-component CF picture can be applied to the $12e+6h$ spectrum
given in Fig.~\ref{fig4}.
The agreement is really quite remarkable and strongly indicates that our 
multi-component CF picture is correct.

In this work we have emphasized that the success of the MFCF picture
is critically dependent on the nature of the pseudopotential.
We have presented several examples of SR pseudopotentials for which
the CF picture works well, and several subharmonic pseudopotentials 
for which it does not.

We gratefully acknowledge partial support from the Materials 
Research Program of Basic Energy Sciences, US Department of Energy.

\end{document}